\documentclass[aps, prd, superscriptaddress, twocolumn, preprintnumbers,nofootnoteinbib]{revtex4-1}
\usepackage[T1]{fontenc}
\usepackage{microtype}
\usepackage[textsize=scriptsize,backgroundcolor=red!70,linecolor=red]{todonotes}
\usepackage{booktabs}
\usepackage{caption}
\usepackage{dcolumn}
\usepackage{amsmath}
\usepackage{amsfonts}
\usepackage{amssymb}
\usepackage{graphicx}
\usepackage{hyperref}
\usepackage[all]{hypcap}
\usepackage{paralist}
\usepackage{multirow}
\newcommand{\nue}{\ensuremath{\nu_e}}
\newcommand{\numu}{\ensuremath{\nu_\mu}}

\newcommand{\chisq}{\ensuremath{\chi^2}}

\newcommand{\dcp}{\ensuremath{\delta_{\text{CP}}}}
\newcommand{\anti}[1]{\ensuremath{\bar{#1}}}

\usepackage{graphicx}
\usepackage{dcolumn}
\usepackage{bm}
\usepackage{epsf}
\usepackage{dcolumn}
\def\be{\begin{equation}}
\def\ee{\end{equation}}
\def\bea{\begin{eqnarray}}
\def\eea{\end{eqnarray}}
\def\gsim{\ \rlap{\raise 2pt\hbox{$>$}}{\lower 2pt \hbox{$\sim$}}\ }
\def\lsim{\ \rlap{\raise 2pt\hbox{$<$}}{\lower 2pt \hbox{$\sim$}}\ }
\def\dslash{\kern-4pt \not{\hbox{\kern-2pt $\partial$}}}
\def\pslash{\not{\hbox{\kern-2pt p}}}

\begin{document}
\DeclareGraphicsExtensions{.eps,.ps}


\title{Configuring the Long-Baseline Neutrino Experiment}



\author{Vernon Barger}
\affiliation{Department of Physics, University of Wisconsin, Madison,
 WI 53706, USA}

\author{Atri Bhattacharya}
\affiliation{
Harish-Chandra Research Institute, Chhatnag Road, Jhunsi,
Allahabad 211 019, India}

\author{Animesh Chatterjee}
\affiliation{
Harish-Chandra Research Institute, Chhatnag Road, Jhunsi,
Allahabad 211 019, India}

\author{Raj Gandhi}
\affiliation{
Harish-Chandra Research Institute, Chhatnag Road, Jhunsi,
Allahabad 211 019, India}

\author{Danny Marfatia}
\affiliation{Department of Physics and Astronomy, University of Hawaii at Manoa,
Honolulu, HI 96822, USA}
\affiliation{Department of Physics and Astronomy, University of Kansas,
Lawrence, KS 66045, USA}
\affiliation{Kavli Institute for Theoretical Physics, University of California, Santa Barbara, CA 93106, USA}

\author{Mehedi Masud}
\affiliation{
Harish-Chandra Research Institute, Chhatnag Road, Jhunsi,
Allahabad 211 019, India}

\begin{abstract}
We study the neutrino oscillation physics performance of the Long-Baseline Neutrino Experiment (LBNE) in various configurations. In particular, we compare the case of a surface detector at the far site augmented by a near detector, to that with 
the far site detector placed deep underground but no near detector. In the latter case, information from atmospheric neutrino events is also utilized. For values of 
$\theta_{13}$ favored by reactor experiments and a 100~kt-yr exposure, we find roughly equivalent sensitivities to the neutrino mass hierarchy, the octant of 
$\theta_{23}$, and to CP violation. We also find that as the  exposure is increased, the near detector helps increase the sensitivity to CP violation substantially more than atmospheric neutrinos.

\end{abstract}
\maketitle

Neutrino oscillations in the standard three flavor framework are 
governed by $a)$  two mass-squared differences 
$\Delta m^2_{j1}=m^2_j-m^2_1$, $j=2,3$, $b)$ three mixing angles
$\theta_{12}$, $\theta_{23}$ and $\theta_{13}$ and $c)$ one Dirac CP phase $\delta_{CP}$.  
The mass-squared differences and the angles $\theta_{12}$ and $\theta_{23}$ are precisely known, and we now have incontestable evidence that 
$\theta_{13} \neq 0$~\cite{db}. 
Answers to outstanding questions, 
such as the nature of the {\it mass hierarchy} ({\it i.e.},
whether the hierarchy is {\it normal} with $m_3>m_1$ or {\it inverted} with $m_3<m_1$), the {\it octant of $\theta_{23}$} 
({\it i.e.}, whether $\theta_{23}$ lies in the lower octant with $\theta_{23}<\pi/4$ or in the higher octant with $\theta_{23}> \pi/4$), and whether CP is violated in the leptonic sector, are now within reach, and will be addressed, for instance, by the Long-Baseline Neutrino Experiment (LBNE)~\cite{LBNE-interim}. 

LBNE is a proposed neutrino detector that will be housed at the Homestake mine. In its first stage, it is envisioned as a 10~kt Liquid Argon (LAr) Time Projection Chamber that will detect neutrinos from a beam produced at Fermilab at a 1300~km baseline. A small near detector (ND)  at Fermilab is a possibility that would assist the LBNE program. 
Detector configurations for LBNE are the subject of much recent discussion~\cite{reconfig}.  
It must be emphasized that both an underground far detector (FD) and the ND have utilities beyond that for oscillation physics. Having the FD underground, 
apart from being sensitive to atmospheric neutrinos, also allows studies of proton decay and the detection and measurement of neutrino fluxes from a core collapse supernova.
A fine-grained tracker as an ND would enable precise measurements of such Standard Model parameters as the weak-mixing angle $\theta_{W}$, and also enable precision measurements of the neutrino nucleon cross sections and searches for new physics. 
In what follows, we only compare the various detector configurations from the point of view of sensitivities to neutrino oscillation physics. 

Placing the FD underground has the advantage that it is sensitive to both beam and atmospheric neutrinos, allowing
the datasets from the two sources to be combined. Since detector magnetization significantly impacts the analysis of atmopsheric neutrinos, where appropriate, we consider both a magnetized and an unmagnetized detetctor. On the other hand, having an  ND is expected to significantly reduce background systematics. We evaluate whether it is better to have an underground FD but no ND, or to have a 
surface FD and an ND. We also provide results for a more ambitious program with an underground FD and an ND.
Specifically,  we present results for the following experiments:
\begin{enumerate}
	\item A beam experiment without an ND with the FD on the surface,
	\item A beam experiment with an ND to help reduce systematics at the surface FD,
	\item An experiment without an ND that combines beam and atmospheric neutrino data collected at the underground FD, and
         \item An experiment with an ND that combines beam and atmospheric neutrino data collected at the underground FD (both magnetized and unmagnetized).
	\end{enumerate}

\begin{figure*}[t]
	\centering
		\includegraphics[scale=0.9]{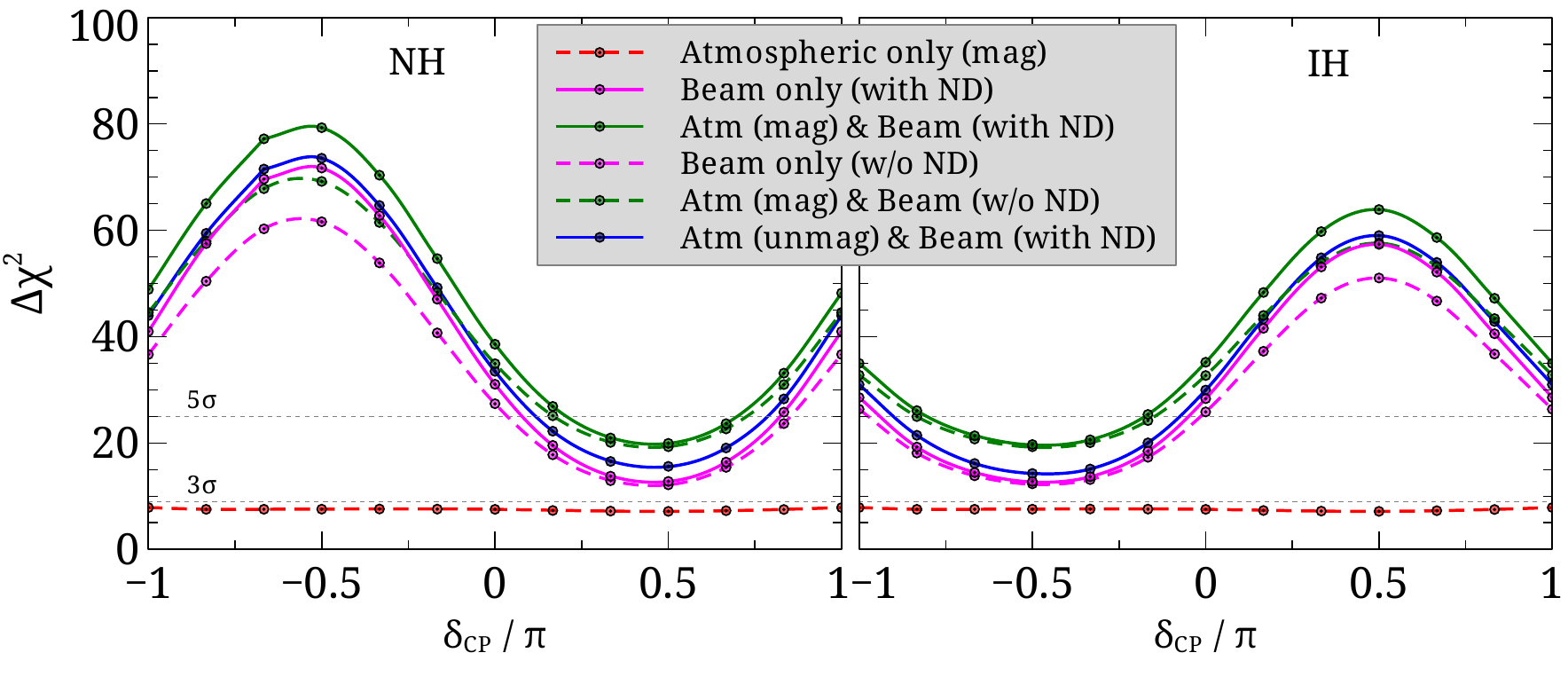}
	\caption{\label{fig:mh-dcp}Sensitivity to the mass hierarchy for a true normal hierarchy (NH) and a true inverted hierarchy (IH) with a 100~kt-yr exposure at the magnetized (mag) far detector configured with and without a near detector (ND), and also with an unmagnetized (unmag) FD and an ND. A run-time of 5~years each with a $\nu$ and $\anti{\nu}$ beam is assumed.}
\end{figure*}

\begin{figure*}[t]
	\centering
	\includegraphics[scale=0.9]{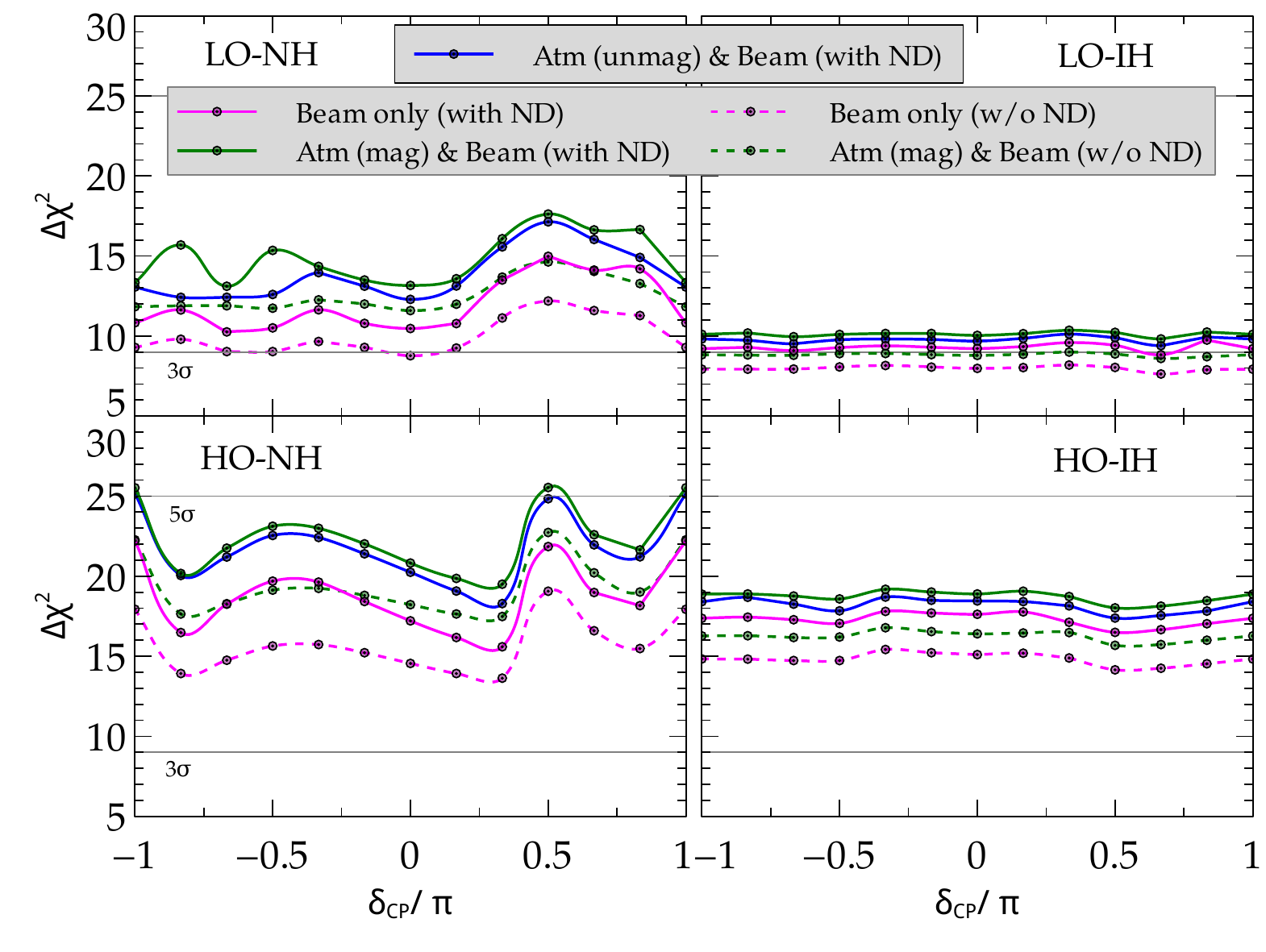}
	\caption{\label{fig:oct-dcp-th13-prior}Sensitivity to the octant with $\sigma(\sin^{2}(2\theta_{13})) = 0.05\times \sin^{2}(2\theta_{13})$, 
	  and $\sin^{2}(\theta_{23}^{\text{true}}) = 0.427$ in the lower octant (LO) and 0.613 in the higher octant (HO), for both hierarchies and a 100 kt-yr magnetized far detector exposure configured with and without a near detector, and also with an unmagnetized FD and an ND.}
\end{figure*}

\begin{figure*}[tbh]
	\centering
		\includegraphics[scale=0.9]{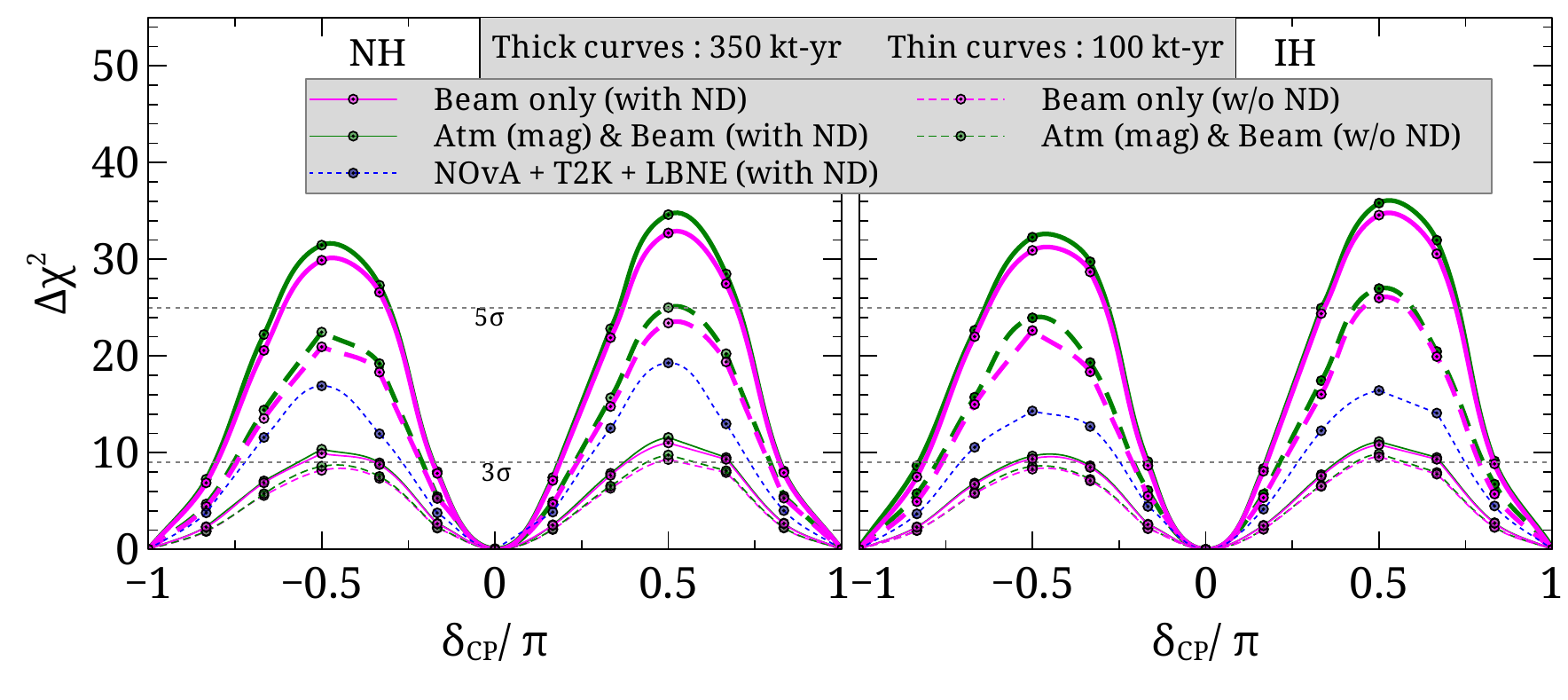}
	\caption{\label{fig:cpv1}Sensitivity to CP violation for 100 kt-yr and 350 kt-yr exposures (with magnetization) assuming $\sigma(\sin^{2}(2\theta_{13})) = 0.05\times\sin^{2}(2\theta_{13})$. Atmospheric neutrino data have very little CP sensitivity. The blue dotted curve shows the combined sensitivity of NO$\nu$A (15~kt TASD, 3 years\ $\nu$ + 
3~years $\bar{\nu}$), T2K (22.5 kt water cherenkov, 5~years\ $\nu$) and LBNE (100 kt-yr LAr FD) beam data.
}
\end{figure*}

Our focus is on a 100 kt-yr exposure that would result from taking data in a 700~kW beam over a 10 year period with 5 years in neutrino mode and 5 years in antineutrino mode. However, we point out and present the effects of a 350~kt-yr exposure in the case of CP violation where the larger exposure makes a  qualitative difference.

LAr detectors are sensitive to \nue\ and \numu\ over the entire energy range of interest, {\it i.e.}, $1-10$~GeV, for beam and atmospheric neutrinos. 
We simulate atmospheric neutrino events and both \nue\ appearance and \numu\ disappearance channel events from the beam in the neutrino and antineutrino modes with an event reconstruction efficiency of 85\%.
For the beam experiment without an ND, we adopt a 5\% systematic uncertainty for the signal (both for \nue\ and \numu) at the far detector, and 10\% and 45\% systematics for the background to the \nue\ and \numu\ signals, respectively. We suppose that the ND reduces systematics at the far detector to 1\% for the \nue\ and \numu\ signals and 5\% and 1\% for the backgrounds, respectively. We employ the signal and background efficiencies, and resolution functions from~\cite{LBNE-interim} and utilize the GLoBES software~\cite{globes}.

For our simulation of atmospheric neutrino data, the energy and angular resolutions of the detector are as described in Ref.~\cite{Barger:2012fx}. We take the atmospheric fluxes from Ref.~\cite{Honda:2004yz}, 
the earth matter density profile from Ref.~\cite{PREM}, and the flux and systematic uncertainties from Ref.~\cite{Barger:2012fx}. The atmospheric neutrino events are divided in energy bins of width 1~GeV from 1--10 GeV.  Events in each energy bin are further subdivided into 18 zenith angle bins in the interval $\cos\theta \in  [-1.0, -0.1]$. 

The atmospheric neutrino data analysis depends on whether the FD is magnetized or not.
For the former case,
$\chi^{2}_{\text{mag}} = \chi^{2}_{\mu^{-}} + \chi^{2}_{\mu^{+}} + \chi^{2}_{e^{-}} + \chi^{2}_{e^{+}}$. 
If the FD is unmagnetized, and charge identification of the produced lepton is not possible, then,
$\chi^{2}_{\text{unmag}} = \chi^{2}_{\mu^{-} + \mu^{+}} + \chi^{2}_{e^{-} + e^{+}}$. 
where $\chi^{2}_{\mu^{-} + \mu^{+}} (\chi^{2}_{e^{-} + e^{+}})$ is the $\chi^{2}$ after summing the $\mu^{-} (e^{-} )$ and $\mu^{+} (e^{+} )$ events.

Throughout, we assume that placing the detector underground does not result in a significant change of the signal and background analysis, apart from making the detector also sensitive to the atmospheric neutrino events, and thus allowing a combined analysis of the beam and atmospheric data simultaneously over the entire duration of the experiment.

To investigate the sensitivity, for instance, to the mass hierarchy using only atmospheric neutrinos, we take the hierarchy to be the normal and evaluate the ability to exclude the inverted hierarchy, and vice versa. We use these events to calculate \chisq~for a fixed set of parameters, assuming the widely used standard
Gaussian approximation, including systematic and statistical uncertainties. We note that the Gaussian approximation has been shown~\cite{lbne_gaussian} to provide an accurate measure of the sensitivity for LBNE and recourse to the  method of Ref.~\cite{qian} is not necessary.  
We then marginalize over the following parameter ranges:
$\theta_{23} \in [0.628, 0.942]$ (i.e., $\left[36^\circ, 54^\circ\right])$, 
$\lvert\Delta m^2_{31}\rvert \in \left[2.19, 2.62\right] \times 10^{-3}\ \text{eV}^2$,
$\theta_{13} \in [0.052, 0.192]$ (i.e., $\left[3^\circ, 11^\circ\right])$,
and
$\dcp \in [-\pi, \pi]$.

%

For the combined analysis of atmospheric and beam events, we sum the \chisq\ obtained from the separate fixed parameter analyses and marginalize over the parameter set, to determine the minimum {\chisq}.


The sensitivities to the mass hierarchy, octant and CP violation are shown in Figs.~\ref{fig:mh-dcp}, ~\ref{fig:oct-dcp-th13-prior} and~\ref{fig:cpv1}.
From these figures, it is clear that, as expected, the best sensitivities 
are obtained from a combination of beam and atmospheric neutrino data collected by an FD in conjunction with an ND (solid green lines in Figs.\ \ref{fig:mh-dcp}--\ref{fig:cpv1}). However, this optimal configuration is not expected for the first phase of LBNE.

As is evident from Fig.~\ref{fig:mh-dcp}, for a range of $\dcp$ values, the determination of the mass hierarchy benefits more from having an underground FD than from having an ND with an FD on the surface (compare the dashed green lines with the solid magenta lines). Specifically,  the underground detector provides significantly better results for the mass hierarchy resolution where the sensitivity to $\dcp$ is generally poor ($\dcp \in [0.25\pi, 0.75\pi]$ for the normal hierarchy and $\dcp \in [-0.75\pi, -0.25\pi]$ for the inverted hierarchy).
%

However, given the recent evidence from reactor experiments that $\theta_{13}$ is quite large~\cite{db}, neither having an ND  nor taking the FD underground is important to  the mass hierarchy determination. Even for the most pessimistic values of \dcp, the mass hierarchy is expected to be resolved at more than the $3\sigma$ confidence level with a 100~kt-yr exposure of a surface FD alone.
Identical conclusions vis-a-vis going underground or having an ND can be drawn for a 35~kt FD with 10 years of data; in this case the mass hierarchy is resolved well beyond $5\sigma$ for all values of \dcp.

It is also worth noting that while atmospheric neutrinos are appreciably less sensitive ($\approx 3\sigma$) to the mass hierarchy when compared to even the poorest performance of the beam experiment, results from atmospheric neutrinos are independent of \dcp~\cite{Gandhi:2007td}.  
In case \dcp\ happens to have a value for which the beam experiment has reduced sensitivity, this feature of atmospheric neutrinos could be an asset for an underground FD.

Resolving the octant degeneracy for the parameter $\theta_{23}$ depends greatly on the precision with which $\theta_{13}$ will be known. For $\sigma(\sin^{2}2\theta_{13}) = 0.01$, for instance, $\Delta \chi^{2}$ decreases from the values shown in Fig.~\ref{fig:oct-dcp-th13-prior} by $30-60\%$ for the various cases.
In addition, as is seen in Fig.\ \ref{fig:oct-dcp-th13-prior}, the combined analysis without an ND (green dashed lines) does marginally better than a beam-only analysis with an ND in place (magenta solid lines) for the normal hierarchy.
For the inverted hierarchy, however, the beam-only results with  an ND are slightly better than those for an underground FD without an ND in place.
These inferences remain unchanged irrespective of whether the FD is magnetized or not, and even if the FD volume is increased to 35~kt.


In the case of CP violation, as Fig.~\ref{fig:cpv1} attests, for a 100~kt-yr exposure neither the presence of an ND nor the inclusion of atmospheric data (for a magnetized FD) markedly improves sensitivity, although the former does slightly better than the latter.  However, the increase in sensitivity due to an ND is significant if the exposure is increased to 350~kt-yr, as can be seen by comparing the thick solid curves with the thick dashed curves in Fig.~\ref{fig:cpv1}.
These  display a $\approx{50\%}$ improvement in $\Delta\chisq$ due to the ND, with significant fractions of the $\dcp$ range above the $3\sigma$ and $5\sigma$ confidence levels. Atmospheric data on the other hand, contribute only a marginal improvement. We have also included a curve showing the significant effect of a joint analysis with NO$\nu$A~\cite{nova} and T2K~\cite{T2K} data.

In summary,  we find that for a 10~kt detector that collects data for 5 years in neutrino mode and 5 years in antineutrino mode, the sensitivities to the mass hierarchy, octant of $\theta_{23}$ and CP, are roughly similar for a beam-only experiment with a near detector, and an experiment without a near detector that combines atmospheric and beam data. For the octant resolution, the latter configuration performs better than the former if the hierarchy is normal, while reverse holds for the inverted hierarchy. The hierarchy can be determined with 100~kt-yr exposure to high significance without assistance from atmospheric data or a near detector.  
Over a 10~year period for a 35 kt~detector, the reduced systematics due to a near detector considerably improves the  CP sensitivity  (since the total uncertainty is  no longer statistics dominated), making it possible to exclude a large fraction of $\dcp$ values at high significance.


 \vspace{0.1 in}
{\it Acknowledgments:} AB, AC, RG and MM  thank Sanjib Mishra and Pomita Ghoshal for very helpful discussions. This work was supported by 
US DOE grants DE-FG02-95ER40896, DE-FG02-13ER42024 and DE-FG02-04ER41308,
and US NSF grant PHY11-25915. RG acknowledges the 
support of the  XI Plan Neutrino Project under DAE  and is grateful to the Phenomenology 
group at the University of Wisconsin-Madison and the LBNE collaboration at Fermilab for its hospitality while this work was in progress.
DM thanks the Kavli Institute for Theoretical Physics for its hospitality during
the completion of this work.

\vskip1cm


\end{document}